# Design Concept of A γ-γ Collider-Based Higgs Factory Driven by a Thin Laser Target and Energy Recovery Linacs

Yuhong Zhang

Thomas Jefferson National Accelerator Facility
12000 Jefferson Avenue, Newport News, VA 23607 USA

***Abstract***   A γ-γ collider has long been considered an option for a Higgs Factory. Such photon colliders usually rely on Compton back-scattering for generating high energy γ photons and further Higgs bosons through γ-γ collisions. The presently existing proposals or design concepts all have chosen a very thick laser target (i.e., high laser photon intensity) for Compton scatterings. In this paper, we present a new design concept of a γ-γ collider utilizing a thin laser target (i.e., relatively low photon density), thus leading to a low electron to γ photon conversion rate. This new concept eliminates most useless and harmful soft γ photons from multiple Compton scattering so the detector background is improved. It also greatly relaxes the requirement of the high peak power of the laser, a significant technical challenge. A high luminosity for such a γ-γ collider can be achieved through an increase of the bunch repetition rate and current of the driven electron beam. Further, a multi-pass recirculating linac could greatly reduce the linac cost and energy recovery is required to reduce the needed RF power.

## 1. Introduction

Since the recent announcement of the discovery of a Higgs-like boson at the Large Hadron Collider (LHC), there is renewed interest [1] in constructing a new accelerator facility, a Higgs Factory, for producing large quantities of these particles for further experimental studies. A Higgs Factory has been studied extensively for a long time, with various options considered; one of them is a γ-γ collider which utilizes the channel γ-γ→$H$.

The concept of a γ-γ collider was first proposed by Russian scientists in 1981 [2] and their extensive work has established the solid theoretical foundation of such a facility [3]. It was initially considered as the second interaction point/detector for the various next generation high-energy linear collider proposals [4,5,6,7,8,9]. Recently, the idea of a stand-alone γ-γ collider as a Higgs Factory, driven by cost efficient recirculating linacs has been proposed [10,11] and at least one set of machine design parameters has been developed [11].

The most favored mechanism of generating hard γ photons for a γ-γ collider is Compton back-scattering, namely, (head-on) scattering of a high intensity laser beam by ultra-relativistic electrons (energy 100+ GeV). The γ photons from the scatterings have a very broad spectrum with a maximum energy about 83% of the electron energy. The conventional approach is to employ a very thick laser target for harvesting a large number of γ photons from Compton scattering in order to achieve a high luminosity. A thick laser target means an extremely high (saturated) laser photon density such that all electrons will be scattered at least once, therefore, generating on average at least one γ photon per electron. While this is generally considered highly efficient, such an approach has several disadvantages including a large number of soft γ photons generated by multiple Compton scattering, and it also requires a very high peak laser power. The soft γ photons have energies much less than half of Higgs mass, thus they don't contribute to generation of Higgs bosons but do increase the detector background.

The new γ-γ collider design concept presented in this paper is based on a thin laser target for Compton scattering. It utilizes a much lower laser photon density such that only a small fraction (typically 20% or less) of electrons are scattered by these laser photons, leading to a much lower electron to γ photon conversion. This ensures the elimination of nearly all the soft γ photons from multiple Compton scatterings, which improves the detector background. On the other hand, one can increase the electron beam current by increasing both the bunch repetition rate and the bunch charge to create a high average flux of γ photons, thus restoring a high luminosity for γ-γ collisions. A higher current electron beam does require a higher RF power to accelerate it, nevertheless, much of that beam power can be recovered due to the fact that a high percentage (up to 80%) of electrons are not scattered at all. Thus, the net RF power consumed in electron to γ photon conversion for this new scheme is still relatively modest. As a matter of fact, it is tied to the luminosity, therefore, resulting in a similar level of power consumption as the thick target design. The additional advantages of this new design concept are that it requires much lower peak laser power and also eliminates the nonlinear effects (including multi-photon scatterings) of strong laser fields.

This tech note is organized as follows: the new design concept will be explained in the next section, followed by a case design for the application of Higgs Factory; in the fourth section, we discuss the energy recovery over multi-pass recirculating linacs.

## 2. The Design Concept

The mechanism for generating γ photons through Compton back-scatterings has been extensively studied; a good review can be found in reference [3]. Below we will use these standard results for presenting the new design concept of a γ-γ collider.

The γ photons from Compton back-scatterings have a broad spectrum, from nearly zero up to a very sharp cut-off at the high energy end. The maximum energy $E_{\gamma,max}$ is [3]

$$\frac{E_{\gamma,max}}{E_e} = \frac{x}{1+x} \qquad (1)$$

where $E_e$ is the electron energy, and the parameter $x$ is defined [3] as

$$x = \frac{4E_e \hbar \omega_0}{(m_e c^2)^2} \approx 15.3 \left[\frac{E_e}{\text{TeV}}\right]\left[\frac{\hbar \omega_0}{\text{eV}}\right] \qquad (2)$$

where $\hbar\omega_0$ is the laser photon energy. It was demonstrated [3] that the optimized value of $x$ is 4.8, resulting in a highest energy $E_{\gamma,max} \approx 0.83 E_e$ for the γ photons. Above this threshold of $x$, the high energy γ photons are likely lost due to $e^+e^-$ pair creations. It has also been shown that, for the case of ultra-relativistic electrons, the γ photon beam from Compton back-scatterings has a similar shape to the electron beam and also follows the same direction of the electron beam [3].

The number of γ photons generated from Compton back-scatterings of one electron bunch depends on the number of laser photons. The electron to γ photon conversion rate is, in the regime of low laser photon density, given by the following formula [3]

$$k = \frac{N_\gamma}{N_e} \sim 1 - e^{-\frac{A}{A_0}} \approx \frac{A}{A_0} \qquad (3)$$

where $A$ is the energy of one laser flash and defined as $A=N_1\hbar\omega_0$, and $N_1$ is number of photons per flash. $A_0$ is a parameter depending on the laser pulse length and Compton cross-section [3]

$$A_0 = \frac{\hbar c l_e}{2\sigma_e} = b l_e [\text{cm}] \text{ J} \tag{4}$$

where $l_e$ is length of the electron bunch and the coefficient $b$ ranges from 8.4 to 25 depending on the pulse length and profile of the laser pulse. It should be noted that formula (3) assumes there is no multiple Compton scattering.

The conventional design strategy for a γ-γ collider is to employ a thick laser target. It is achieved by pushing up the laser flash energy such that $A$ is equal to or large than $A_0$. A "standard" design calls for $A=A_0$, thus the electron to γ photon conversion rate is estimated at about 63%, under the assumption that the formula (the left part) in Eq. (3) is still valid. There are several design proposals in which the laser flash energy are even higher, such that the electron to γ photon conversion rates are, for examples, as high as 2.4 and 1.2 for CLICHÉ [8] and SAPPHiRE [10], respectively.

Nevertheless, there are several issues associated to the design approach with thick laser targets. Below we will briefly discuss some of these issues and point out how they affect the collider performance or pose serious technological challenges.

The first issue is a dramatic increase in the probability of multiple Compton scatterings of an electron, generating a large amount of low energy soft γ photons, as illustrated in Figure 1. The inserted table in the left plot lists the ratio of total γ photons ($N_\gamma$) to those generated by the electrons having only a single scattering ($N_{\gamma,\text{single}}$) as a function of thickness of the laser target. The data ($N_\gamma/N_{\gamma,\text{single}}$) is plotted in the middle figure, showing clearly these soft γ photons from multiple Compton scatterings increase almost linearly proportional to the laser flash energy $A$. The right plot shows the same data again but in terms of $N_{\gamma,\text{single}}/N_\gamma$, namely, γ photons generated in single Compton scatterings as a percentage of the total γ photons. It shows that the ratio drops rapidly as the laser target thickness increases, clearly demonstrating that a thick laser target is ineffective for a Higgs Factory. As an example, one can find that, when $A=A_0$, more than half of γ photons are generated by secondary Compton scatterings. At $A=2A_0$, nearly 80% of γ photons are soft ones.

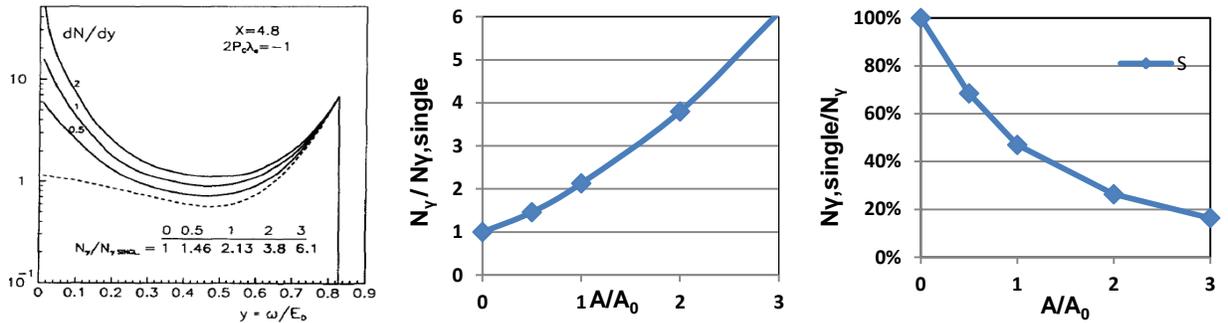

Figure 1: (Left) Normalized (to maximum) photon spectra for different numbers of interaction lengths in a laser target (these numbers mark the solid curves); dashed curve is photon spectra without secondary scatterings (adopted from Figure 4 of reference [10]); (Middle) The ratio of total γ photons over those generated by single Compton scatterings; (Right) The same data plotted in percentage the γ photons from single Compton scatterings over the total γ photons.

The second issue is the ultra high peak laser power required to support a saturated photon density. A thick laser target demands a laser flash energy A ≥ $A_0$, and a typical value of $A_0$ is 1 J. With a typical 0.1 mm RMS electron bunch length for an electron beam from a linac, the peak laser power is in an order of 1 TW. On the other hand, the average laser power is fairly modest, in the range of several tens kW, for either pulsed beams or CW beams with a low bunch repetition rate.

The third issue is the influence of a strong electromagnetic field of the laser on the electron to γ photon conversion. When the laser photon density is very high, the multi-photon effects such as e+nγ→e+nγ$_0$ and e+nγ→e$^+$e$^-$ become non-negligible, where γ and γ$_0$ denotes laser photons and n is the number of laser photons involved in the scatterings. Such nonlinear effects are characterized by the following parameter [3]

$$\xi^2 = \left[\frac{4F\hbar}{m_e \omega_0 c}\right]^2 \approx \frac{2}{\pi a} \frac{\sigma_c}{\sigma_0} \frac{\lambda}{l_\gamma} k \tag{5}$$

where $F$ represents the $E$ or $B$ field of the laser, $a$ is spot size of the electron beam at the collision point, $l_\gamma$ is the laser pulse length, $\sigma_c$ and $\sigma_0$ are Compton cross-sections associated to $x=0$ (non-relativistic) and $x\neq 0$ respectively, and $\sigma_c/\sigma_0\sim 0.75$ when $x=4.8$. The parameter $k=N_\gamma/N_e$ is the electron to γ conversion rate, and is determined by the laser photon density. The general rule is that when $\xi^2\ll 1$, the nonlinear field effect of the high intensity laser beam is negligible. Otherwise, two or more laser photons can be scattered at the same time. There are several bad effects of these nonlinear fields, one of them is the reduction of the maximum energy of γ photons according to the following formula [6]

$$\frac{E^n_{\gamma,max}}{E_e} = \frac{nx}{1+\xi^2+nx} \tag{6}$$

where $n$ is number of laser photons involved in the scatterings. When $\xi^2\ll 1$, Eq. (6) at $n=1$ is identical to Eq. (1). If, on the other hand, $\xi^2\sim 1$, the maximum energy of γ photons from single Compton scatterings ($n=1$), is reduced to 71% of the electron energy for $x=4.8$, a significant drop from the optimized value of 83% $E_e$.

The new design concept for a γ-γ collider aims to address the issues associated with the thick laser target design approach. The key feature of this new concept is a thin laser target, namely, $A/A_0 \ll 1$, therefore, it could

- eliminate effectively all the soft γ photons generated by multiple Compton scatterings;
- lower the peak laser power by the same reduction factor of $k$;
- reduce the parameter $\xi^2$ proportionally, thus significantly reducing the nonlinear effect of the high intensity laser beam.

It is obvious that reduction of the laser target thickness leads to less γ photons, hard and soft, from Compton back-scatterings, thus resulting in a lower luminosity. However, one can increase the electron beam current through higher bunch repetition rate and higher bunch charge to compensate for the luminosity loss. Specifically, for the broadband luminosity, assuming the same electron beam parameters and interaction region design,

$$\frac{L_{b,thin}}{L_{b,thick}} = \frac{f_{thin}}{f_{thick}} \left[\frac{N_{e,thin}}{N_{e,thick}}\right]^2 \left[\frac{k_{thin}}{k_{,thick}}\right]^2 \tag{7}$$

where $L_{b,thin}$ $L_{b,thick}$, $f_{thin}$, $f_{thick}$, $N_{e,thin}$, $N_{e,thick}$, $k_{e,thin}$ and $k_{e,thick}$ are the broadband luminosities, bunch repetition rates, numbers of electrons per bunch and the electron to γ photon conversion rates for the thin and thick laser targets respectively. It is not difficult to achieve $L_{b,thin}/L_{b,thick}$~1 or even larger. As an example, if $f_{thin}/f_{thick}$=5, $N_{e,thin}/N_{e,thick}$=2, and $k_{e,thin}/k_{e,thick}$=0.14/0.63~0.22, then $L_{b,thin}/L_{b,thick}$~1. Though the bunch charge is doubled in this example, the laser target thickness is nevertheless reduced by a factor of 1/0.15≈6.7, therefore, the number of the laser photons per flash is reduced by a factor of 3.3 between the thin and thick target designs.

A higher current electron beam requires a higher RF power to accelerate it to the working energy. Since the beam energy is very high, from 80 to 500+ GeV, the required increase of the average RF power is very significant. However, since most of the electrons are not scattered by the laser photons at the conversion area, their energy could be recovered in the same SRF linacs that accelerates them. Presently, the energy recovery technology is well developed such that a full energy recovery could be expected. This technical topic will be discussed in section 4.

## 3. Design Parameters

In the following, we present the main parameters for an exemplary γ-γ collider based Higgs Factory using Compton back-scattering with a thin laser target. This new design is, in several aspects, similar to SAPPHiRE [11]. The differences of the two designs, beside different laser target thicknesses, are

- the parameter $x$ is shifted to the optimized value 4.8;
- the electron energy is increased to 100 GeV;
- the RMS bunch length of the electron beam is 0.1 mm.

We argue that a higher electron beam energy would ensure that a γ photon with about 60% of that electron energy is still above half of the mass of the Higgs-like boson (125 to 127 GeV as the latest best experimental value), thus contributing to the generation of such new particles. On the contrary, the energy range of γ photons that could contribute to the generation of Higgs-like boson in SAPPHiRE is very narrow (62.6 to 64.9 GeV), therefore, it is fairly ineffective.

Table 1 lists the key parameters for the collider design. The laser flash energy is chosen such that $A/A_0$=0.15, thus the electron to γ photon conversion rate is about 14%. This value is purely empirical and for the purpose of illustration. An optimized value could be achieved through additional studies and supported by simulations. As a comparison, we also list the parameters for the SAPPHiRE design [11].

The broadband γ-γ collider luminosity in Table 1 is obtained through a simple formula $L_{\gamma\gamma}=k^2 L_{e^-e^-}$ where $L_{e^-e^-}$ is the geometric luminosity for $e^-e^-$ collision and $k$ is the electron to γ photon conversion rate. The γ-γ collider luminosity with γ photon energy above 60% of the electron energy is estimated by multiplying another reduction factor, $(k')^2$, where $k'$ is the percentage of the hard γ photons (i.e., $E_\gamma$>60%$E_e$). We choose $2\lambda P_c$=-1 where λ and $P_c$ are the helicities of the polarized electrons and laser photons. It can be estimated, using Figure 4 of reference [3], that k'~50%, thus $L_{\gamma\gamma}(E_\gamma>60\%E_e)=0.5^2 k^2 L_{e^-e^-}$. This result is also consistent with the estimation using the spectrum luminosity plot in Figure 7 of the same reference [3].

As shown in Table 1, the thin laser target reduces the parameter $\xi^2$ to 0.04, an extremely small value, ensuring there is no nonlinear effect of the laser beam fields.

Table 1: The design parameters for γ-γ collider based Higgs Factory

|  |  | SAPPHiRE* | Thin-Target |
|---|---|---|---|
| Electron energy | GeV | 80 | 100 |
| Electron beam polarization |  | 80% | 80% |
| Electron beam current | mA | 0.32 | 2.4 |
| Bunch repetition | MHz | 0.2 | 1 |
| Electrons per bunch | $10^{10}$ | 1 | 1.5 |
| Electron bunch charge | nC | 1.6 | 2.4 |
| Electron bunch length, RMS | μm | 30 | 100 |
| Normalized emittance, horizontal & vertical | μm | 5 & 0.5 | 5 & 0.5 |
| Beta function at IP, horizontal & vertical | mm | 5 & 0.1 | 5 & 0.1 |
| Electron beam spot size at IP, hori. & vert., RMS | nm | 400 & 18 | 357 & 16 |
| Distance between IP and CP | mm | 1 | 2 |
| Electron beam spot size at CP, hori. & vert., RMS | nm | 154 & 131 | 385 & 320 |
| Crab crossing angle (in horizontal plane) | mrad | 20 | 20 |
| e⁻e⁻ geometric luminosity | $10^{34}$ cm$^{-2}$s$^{-1}$ | 2.2 | 31.3 |
| Parameter $x$ |  | 4.3 | 4.8 |
| Electron to γ photon conversation rate ($k=N_\gamma/N_e$) |  | 1.2 | 0.14 |
| Number of γ photons per electron bunch ($N_\gamma$) | $10^{10}$ | 1.2 | 0.21 |
| Nonlinear effect parameter ($\xi^2$) |  |  | 0.04 |
| γ-γ luminosity (broadband) | $10^{33}$ cm$^{-2}$s$^{-1}$ |  | 6.1 |
| Reduction factor $k'$ for $E_\gamma>0.6E_0$ per beam |  |  | ~0.5 |
| γ-γ luminosity ($E_\gamma>0.6E_0$) | $10^{33}$ cm$^{-2}$s$^{-1}$ | 3.6 | 1.5 |

\* The parameters for SAPPHiRE are from reference [11].

The following table summarizes the laser parameters for SAPPHiRE and the thin laser target design. It can be seen that the peak laser power and peak photon density are reduced significantly for the thin laser target design.

Table 2: Laser parameters for a γ-γ collider based Higgs Factory

|  |  | SAPPHiRE* | Thin-Target |
|---|---|---|---|
| Laser wavelength ($\lambda$) | μm | 0.351 | 0.395 |
| Laser photon energy ($\hbar\omega$) | eV | 3.53 | 3.14 |
| Laser target parameter ($A_0$) | J |  | 0.5 |
| Laser flash energy ($A$) | J |  | 0.075 |
| Laser pulse length ($l_\gamma$) | mm |  | 0.2 |
| Laser peak power | GW |  | 112 |
| Laser peak intensity | W/cm$^2$ | 2.96·10$^{20}$ | 5.7·10$^{17}$ |
| Laser photons per flash | $10^{17}$ |  | 1.5 |
| Laser photons per electron | $10^7$ |  | 1 |
| Laser peak photon density | photons/cm$^2$/s | 1.1·10$^{40}$ | 1.1·10$^{36}$ |
| Laser repetition rate | MHz | 0.2 | 1 |
| Laser average power | kW |  | 75 |

\* The parameters for SAPPHiRE are from reference [11]

In both Table 1 and 2, the design parameters for SAPPHiRE are quoted from the original paper [11]. There could be inconsistencies between the two parameters sets due to using different formulas.

## 4. Recirculating Linac and Energy Recovery

We now consider electron beam acceleration in a γ-γ collider. For the thin laser target design presented above, a total of 480 MW average RF power is needed to accelerate two 2.4 mA average current electron beams to 100 GeV. Such a high RF power requirement is likely to be prohibitive in most accelerator proposals. Fortunately, much of the beam energy in this thin laser target design can be recovered due to the fact that approximately 86% of electrons are never scattered by laser photons. It can be shown that with energy recovery, the net RF power consumption is comparable to that of the conventional thick target designs. Table 3 shows the RF power budgets for both SAPPHiRE and the thin laser target design. We divide the electron beam of the thin laser target design to two parts: a used beam of 0.334 mA and an unused beam of 2.066 mA, according to whether the electron has been scattered by laser photons or not. At the moment, we propose to only recover the energy of the unused beam, thus saving about 413 MW of RF power, while the used beam is simply dumped as in all other γ-γ collider proposals. Since the γ photons have a broad spectrum, it may be possible to recover at least part of the energy of the used beam, thus further reducing the RF power consumption. Such an idea is currently under active evaluation.

Table 3: The RF power budgets for a γ-γ collider

|  |  | SAPPHiRE | Thin-Target |
|---|---|---|---|
| Electron energy | GeV | 80 | 100 |
| Electron current | mA | 0.32 | 2.4 |
| e⁻→γ conversion rate |  | 120% | 14% |
| Used beam | mA | 0.32 | 0.334 |
| Unused beam | mA | 0 | 2.066 |
| Total beam power | MW | 51.2 | 480 |
| Used beam power | MW | 51.2 | 66.9 |
| Unused beam power | MW | 0 | 413.1 |

\* The parameters for SAPPHiRE are from reference [11]

The simplest energy recovery scheme for a γ-γ collider is based on the so-called "push-pull" concept [12,13] as shown schematically in Figure 2. It looks very similar to a conventional linear collider with the two electron to γ photon conversion points (CP) and an γ-γ interaction point (IP) at the center. Each of the two electron beams is accelerated in one SRF linac and then passes through the CP-IP-CP complex. The two beams go straight into the SRF linacs on the other side respectively to have their energy recovered for accelerating the next bunches.

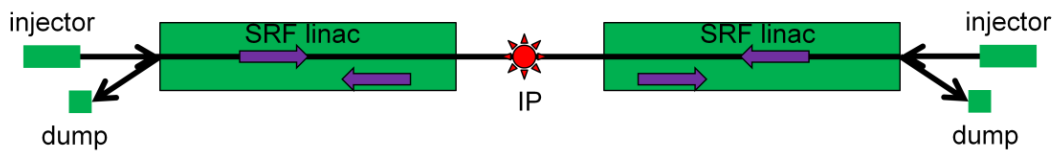

Figure 2: A drawing of "push-pull" energy recovery scheme

Inspired by the success of the Jefferson Lab CEBAF facility, single or multi pass recirculating SRF linacs become a very attractive cost efficient option for accelerating electrons or other charge particles to high or very high energy for physics experiments and for light source applications. Currently there are about a half dozen design studies for recirculating linac applications such as electron-ion collider (LHeC and eRHIC), muon collider and linac-driven 4th generation light source (JLAMP), some of them also combined with energy recovery. This recirculating linac option has also been considered for [10] or incorporated into [11] the γ-γ collider designs. A more recent study [14] examined various design options (geometry/topology and number of passes) for accelerating an election beam to 80 GeV by recirculating linacs, for the purpose of reaching design optimization in terms of linac and beamline/tunnel costs, synchrotron radiation loss and beam quality preservation. In the following we select the most simple and also highly efficient recirculating linac scheme to complete our study. It is straightforward to repeat this study employing other recirculating linac design options.

The schematic drawing of the recirculating linacs with energy recovery for a γ-γ collider is shown in Figure 3. It is basically a linear collider with a recirculating pass consisting of two 180° arcs and a straight beamline. The two SRF linacs are set on both sides of the IP, like the "push-poll" scheme in Figure 1. However, the electron beam injection points are now next to the interaction point in order to let electron beams being accelerated in both linacs before reaching the conversion points. After passing the CP-IP-CP complex, the electron beams take exactly the same number of recirculation passes for energy recovery. Table 4 shows the parameters for the recirculating linacs and also the synchrotron radiation loss in arcs for both accelerating and deceleration (for the unused beam only, assuming the used beam is deflected, collected and sent to a dump by the dipole magnets near the IP). We also include the "push-pull" design in the table. We performed calculations assuming a 1 km bending radius for the recirculation ring of 1.5 km arc radius.

It is clear that, for the case of high electron current required for the thin laser target γ-γ collider design, the synchrotron radiation loss is very significant, thus options of three or more passes of the SRF linacs are out of the question. Only "push-pull" (which has no synchrotron radiation loss) and one or two passes should be considered and one of them could be an optimized solution in terms of balances of linac cost and RF cost (both hardware and operation). Beam quality preservation in the recirculating linacs is also an important issue to be considered in selecting the baseline design. It is well understood that a high energy electron beam experiences serious emittance and energy spread degradation due to strong synchrotron radiation in the recirculating arcs. A detailed study of this issue can be found in reference [13].

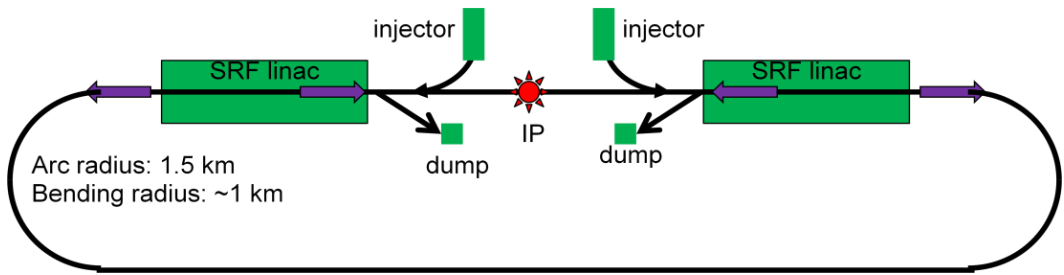

Figure 3: A recirculating linac scheme for a γ-γ collider.

Table 4: Parameters of recirculating linacs for a γ-γ collider

| Recirculating pass | | "push-pull" | 1 | 2 | 3 |
|---|---|---|---|---|---|
| Single linac | GeV | 100 | 50 | 25 | 16.7 |
| Total installed linac | GeV | 200 | 100 | 50 | 33.3 |
| Linac reduction factor | | | 50% | 25% | 16.7% |
| Total SR loss (up or down) | GeV | 0 | 0.55 | 2.84 | 4.83 |
| SR power (up, for 2.4 mA) | MW | 0 | 1.32 | 6.81 | 11.6 |
| SR power (down, for 2.07 mA) | MW | 0 | 1.14 | 5.86 | 9.98 |
| SR power (ERL=up+down) for one beam | MW | 0 | 2.47 | 12.7 | 21.6 |
| for two beam | MW | 0 | 4.94 | 25.3 | 43.2 |
| Total RF power (γ photon + SR loss) | MW | 66.9 | 71.8 | 92.2 | 110 |

## 5. Summary and Discussions

In this paper we presented a new design concept for a γ-γ collider based Higgs Factory. The design concept is based on a thin laser target (relatively low laser photon density) for Compton back-scattering. The advantages of this new design concept are: eliminating all soft γ photons created by multiple Compton scattering for a better detector background; a much lower peak laser power for less accelerator technology R&D; and ensuring no nonlinear effects of the laser fields. Recovery of the energy of the unused (un-scattered) electron beam is a necessary ingredient for this design concept to lower the required RF power. The paper also examines the options of recirculating linacs with energy recovery.

The new γ-γ collider design concept must be validated with further studies. Several issues not addressed in this paper must be included in these studies. We list some of these issues below. The first issue is impact of the new design concept on the detector. There are several aspects of this issue including detector compatibility with a higher bunch repetition rate and much higher $e^-e^-$ collisions. The second issue is the impact on the detector background. While the new design concept greatly reduces soft γ photons by eliminating these generated by all multiple Compton scattering, it is expected the beamstrahlung will become much stronger due to the higher electron beam current. This effect on the detector background must be carefully examined. The third issue is energy recovery. Though it is expected that this accelerator technology should work in the range of energy and current of interest to a γ-γ collider based Higgs Factory, further studies must focus on dynamics and instabilities in such SRF linac systems, particularly when we proceed to recover the energy of some lightly scattered electrons (i.e., those electrons that lose a very small amount of energy in Compton scattering).

The last comment we would like to add is about the optimization of the parameters for Compton back-scattering. The laser target thickness (in terms of $A/A_0$), or more specifically, the laser photon density, largely determines the outcome of the Compton back-scattering (in terms of total γ photons and the ratio of soft to hard γ photons), and further the γ-γ collider performances (luminosity and detector background). It will be very useful to develop clear guidelines [15] for examining and quantifying any additional underlining physics mechanisms. It is expected that during the technical design stage of such a facility, computer simulations will be definitely required for the fine-tuning this parameter and others as well.

## Acknowledgement

The author would like to thank Dr. Alexander W. Chao of SLAC for his encouragement. The author also would like to thank his Jefferson Lab colleagues, Drs. Andrew Hutton, Rui Li and Edward Nissen, for helpful discussions.